\documentclass[manuscript]{acmart}
\acmJournal{TACO}
\setcopyright{none}

\usepackage{subcaption}
\usepackage{bm}
\usepackage{makecell}
\usepackage{listings}
\usepackage[rflt]{floatflt}
\usepackage{multicol}
\usepackage{paracol}

\makeatletter
\AtBeginDocument{\begingroup
  \normalsize
  \let\tmp@n@s\f@size
  \let\tmp@n@b\f@baselineskip
  \small
  \let\tmp@s@s\f@size
  \let\tmp@s@b\f@baselineskip
  \xdef\semismall@size{\fpeval{(\tmp@n@s+\tmp@s@s)/2.1}}\xdef\semismall@baselineskip{\fpeval{(\tmp@n@b+\tmp@n@b)/2.1}}\xdef\semicaptionsmall@size{\fpeval{(\tmp@n@s+\tmp@s@s)/2.25}}\xdef\semicaptionsmall@baselineskip{\fpeval{(\tmp@n@b+\tmp@n@b)/2.25}}\endgroup
}
\newcommand{\semismall}{\fontsize{\semismall@size}{\semismall@baselineskip}\selectfont}
\newcommand{\code}[1]{{\semismall{\texttt{#1}}}}
\newcommand{\smallcode}[1]{{\small{\texttt{#1}}}}
\newcommand{\semicaptionsmall}{\fontsize{\semicaptionsmall@size}{\semicaptionsmall@baselineskip}\selectfont}
\newcommand{\captioncode}[1]{{\semicaptionsmall{\texttt{#1}}}}
\newcommand{\changed}[1]{#1}

\newcommand{\highlight}[1]{\textcolor[RGB]{27, 96, 207}{\bm{#1}}}

\begin{document}

\title{Equality Saturation for Optimizing High-Level Julia IR}
\titlenote{New Paper, Not an Extension of a Conference Paper.}
\author{Jules Merckx}
\orcid{0009-0004-9565-4046}
\email{jules.merckx@ugent.be}
\affiliation{\institution{Computing Systems Lab, Ghent University}
\city{Ghent}
    \country{Belgium}
}

\author{Tim Besard}
\email{tim@juliabhub.com}
\orcid{0000-0001-7826-8021}
\affiliation{\institution{JuliaHub}
\city{Cambridge, MA}
\country{USA}
}

\author{Bjorn De Sutter}
\orcid{0000-0003-0317-2089}
\email{bjorn.desutter@ugent.be}
\affiliation{\institution{Computing Systems Lab, Ghent University}
\city{Ghent}
    \country{Belgium}
}

\begin{abstract}
Compilers are indispensable for transforming code written in high-level languages into performant machine code, but their general-purpose optimizations sometimes fall short.
Domain experts might be aware of optimizations that the compiler is unable to apply or that are only valid in a particular domain.
We have developed a system that allows domain experts to express rewrite rules to optimize code in the Julia programming language.
Our system builds on e-graphs and equality saturation. It can apply optimizations in the presence of control flow and side effects.
As Julia uses multiple dispatch, we allow users to constrain rewrite rules by argument types, and propagate type information through the e-graph representation.
We propose an ILP formulation for optimal e-graph extraction that exploits opportunities for code reuse and introduce \emph{CFG skeleton relaxation} to rewrite calls to pure functions as well as those with side effects.
Use cases demonstrate that our system can perform rewrites on high-level, domain-specific code, as well as on lower-level code such as Julia's broadcasting mechanism. We analyze the required compilation time \changed{and the performance impact of these rewrites.}
\end{abstract}



\setcopyright{none}

\widowpenalty=10000
\clubpenalty=10000
\brokenpenalty=10000

\maketitle

\newcounter{jmcnt}
\newcounter{bdscnt}

\newcommand{\jm}[1]{\refstepcounter{jmcnt}
	\textcolor{brown}{\textbf{JM [\thejmcnt]:} #1}}
\newcommand{\bds}[1]{\refstepcounter{bdscnt}
	\textcolor{purple}{\textbf{Bjorn [\thebdscnt]:} #1}}
	
\section{Introduction}
Optimizing compilers have become indispensable as programming languages rely on them to unravel high-level abstractions into lean and performant code~\cite{lattner2020mlir,hosteColeCompilerOptimization2008,lattnerLLVMCompilationFramework2004}.
General-purpose compiler middle-ends typically perform general-purpose optimizations such as common subexpression elimination~\cite{10.5555/286076}. Compilers for domain-specific languages perform domain-specific optimizations such as algebraic optimizations~\cite{lattner2020mlir,10.5555/286076}. The front-ends of higher-level language compilers can be customized to generate IR that is already optimized for specific targets, such as GPUs, and then fed to the back-ends~\cite{besardEffectiveExtensibleProgramming2019}. Many compilers can also be configured to enable context-specific optimizations, e.g., to sacrifice precision for performance.  
However, none of those existing optimization pipelines are well suited to support domain-specific and context-specific optimization of code written in a high-level scientific programming language that is used across many domains, such as Julia~\cite{bezansonJuliaFastDynamic2012}.
In such a language, software libraries build on the generic language infrastructure to provide high-level domain-specific APIs to application developers. 

Ideally, those application developers should be able to enjoy domain-specific optimizations for free, i.e., without having to manually tune or rewrite their code, and without facing restrictions on their freedom to exploit the language's rapid prototyping features. Furthermore, it should be easy for application developers to specify or select the context-specific optimization they want enabled. In addition, when multiple libraries are being reused and composed in an application, be it top-level libraries that provide orthogonal functionality or higher-abstraction-level libraries that provide more abstract APIs on top of lower-level code, the optimization opportunities coming with those libraries should be composable. They should be composable with each other and with the general-purpose and target-specific optimizations that are provided in the main compiler flow and in the target-specific extensions thereof. 
This inevitably involves solving a phase-ordering problem because some optimizations create opportunities for further optimizations, while others are incompatible and rule each other out.
Finally, to obtain a thriving ecosystem of domain libraries, the development of such libraries and the corresponding support for domain-specific optimization should not require the involvement of compiler experts. 
It then follows that a generic compiler infrastructure is needed that enables developers of domain-specific libraries to specify domain-specific and context-specific optimization opportunities at an abstraction level similar to that of their libraries' APIs, and that ensures sufficient composability. 

With our research, we aim to provide that generic compiler infrastructure. In this paper, we present a system that allows programmers, not necessarily compiler developers, to express their domain-specific knowledge as rewrite rules in Julia.
Our work is based in part on ideas developed for the Cranelift compiler~\cite{fallinAegraphsAcyclicEgraphs2023}, which in turn uses e-graphs and equality saturation, first introduced by \citeauthor{nelsonTechniquesProgramVerification1979}~\cite{nelsonTechniquesProgramVerification1979} and \citeauthor{tateEqualitySaturationNew2009}~\cite{tateEqualitySaturationNew2009}, respectively.
Equality saturation, or EqSat, is a rewrite technique that elegantly deals with the phase-ordering problem by representing the potential results of rewrites while still representing the original program as well.

When writing and executing code, users should not take these rewrite rules into account anymore, but instead can focus on writing simple and readable code that maps well onto their mental model of the computations. The compiler then automatically evaluates all relevant rewrites and picks the optimal program. By building on e-graphs and EqSat, the phase-ordering problem is no longer an issue.
Rewrite rules are applied without removing information from the e-graph, which means that application of a rewrite cannot prevent other rules from firing.

The contributions in this paper are as follows: 
(i) the first use of EqSat in a high-level, dynamically typed programming language optimizer, including for code containing control flow;
(ii) \textit{CFG skeleton relaxation} to enable rewriting on functions with side effects; 
(iii) a novel ILP formulation to enable value reuse in (rewritten) expressions; 
(iv) a novel e-class analysis to track type information for type-constrained rewrite rules;
(v) a demonstration on a number of examples; 
(iv) an analysis of the required compilation time;
\changed{(vi) an analysis of the potential performance gains achievable}.

This paper is structured as follows. Section~\ref{sec:background} provides background. Section~\ref{sec:rewriter} presents our rewriting system. Section~\ref{sec:usecase} presents its capabilities on use cases. Section~\ref{sec:compiletime} analyzes its compilation time, \changed{and Section~\ref{sec:performance} demonstrates obtainable performance optimizations.} Sections ~\ref{sec:limitations} and~\ref{sec:related work} discuss limitations and related work. Finally, Section~\ref{sec:conclusions} draws conclusions.

\section{Background}
\label{sec:background}

\begin{figure}
\centering
    \includegraphics[width=5cm]{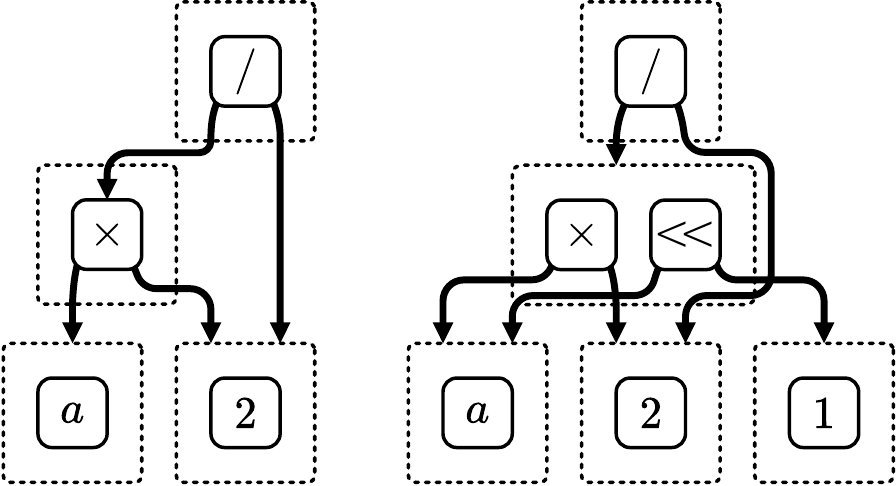}
    \Description{On the right an e-graph with four e-class nodes is shown for the expression a times two divided by 2: one e-class for each operator and one for each distinct operand values. On the left, an extension of the e-graph is shown in which the equivalent of a times 2 in the form of a shifted to the left by one is included. This graph has two more nodes for the extra shift operation and the extra operand value of 1, but the e-nodes for the times and shift operations form one e-class because of the equivalence.}
    \caption{Left: original e-graph~\cite{willseyEggFastExtensible2021} representing the term $(a \times 2) / 2$. Right: e-graph after introducing the equivalence $a \times 2 \leftrightarrow a << 1$.}
    \label{fig:e-graph}
\end{figure}

\emph{E-Graphs.} An e-graph compactly represents a congruence relation on different expression trees~\cite{nelsonTechniquesProgramVerification1979,willseyEggFastExtensible2021}.
It consists of a collection of \textit{e-classes} that can each contain multiple \textit{e-nodes} representing expression trees that are equivalent to each other according to a user-defined equivalence relation.
Each e-node can have multiple children, represented by e-classes.
As an example, the left side of Fig.~\ref{fig:e-graph} shows the e-graph for the expression $(a \times 2) / 2$.
In case $a$ represents an integer, a potential equivalent expression for $a \times 2$ is $a << 1$.
We can encode this equivalence by adding a new e-node to the appropriate e-class, as shown on the right side of Fig.~\ref{fig:e-graph}. Because the constant expression 1 is not equivalent to any of the e-classes already present, it is added to a newly created e-class.
The strength of e-graphs lies in the fact that both equivalent representations are encoded in the graph at the same time.

\paragraph{Equality Saturation}
Besides in theorem solvers~\cite{demouraEfficientEMatchingSMT2007,nelsonTechniquesProgramVerification1979,demouraZ3EfficientSMT2008}, e-graphs are often used for EqSat~\cite{willseyEggFastExtensible2021,cheliMetatheoryjlFastElegant2021,tateEqualitySaturationNew2009}.
EqSat is a term rewriting technique that applies rewrites not by overwriting the original term but by adding the rewritten term to an e-class that represents the original term.
By building on e-graphs for term rewriting, it is possible to represent a much larger number of rewritten, equivalent expressions without the exponential explosion that can occur with naive approaches.
The advantage of not discarding expressions when they get rewritten is that the ordering in which rewrites are applied cannot prevent certain rewrites from firing.
\changed{Central to EqSat is the rebuilding procedure, which is responsible for maintaining the e-graph invariants. When rewriting uncovers two different expressions to be equivalent, all expressions containing these equivalent expressions as subexpressions need to be verified, as these potentially also need to become equivalent now.}

\paragraph{Extraction}
Once no more new rules can be applied or some timeout is reached during EqSat, one typically wants to determine the optimal expression contained in the e-graph~\cite{willseyEggFastExtensible2021,goharshadyFastOptimalExtraction2024}.
In its simplest form, extraction starts from a root e-class where one e-node needs to be picked to be extracted. Recursively, all the child e-classes of the picked e-node need to be extracted as well.
The end result is a directed, connected graph of e-nodes.

In many use cases, especially in program optimization, the extracted graph cannot contain cycles.
In such cases, more refined extraction techniques are needed.
Ours will be discussed in Section~\ref{sec:extraction}.

\paragraph{Julia} Julia is a general-purpose programming language that is primarily used for scientific computing. \changed{Being a dynamically typed language, functions can be defined for unconstrained argument types. A function can also be redefined for specific argument types (and counts) to express that it should behave differently for that combination of types. The different behaviors, i.e., the definitions of a function for different types and counts of arguments, are called \emph{methods} in Julia. In essence, a function is an identifier, such as \code{f} or \code{print}, that it shares with all its methods. When a function call is to be executed, the Julia run-time system will look up and dispatch the function's method that best matches the types and count of all the call's run-time arguments. This is called \emph{multiple dispatch}.

In Julia, operators are syntactic sugar for functions. Even basic functionality like array indexing can be customized with methods for different types of arrays and indices. A method can, e.g., redefine how elements in a \code{Symmetric} matrix are accessed, avoiding the need to store symmetric elements twice. Methods that implement more abstract computations on matrices by accessing those matrices' elements through index operations will automatically benefit from that custom access method when they are invoked on a \code{Symmetric} matrix. \changed{Because all operators are functions, Julia expressions correspond by and large to trees of function calls.}
}

\begin{figure}
\centering
\includegraphics[width=0.47\linewidth]{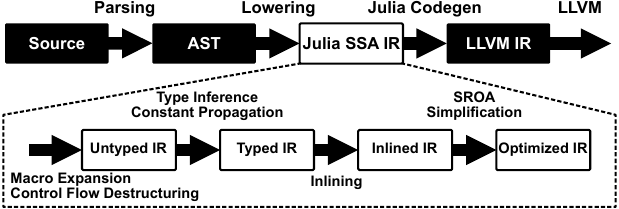}
    \caption{The native Julia compilation pipeline.}
    \Description{On top, the Julia compiler pipeline is shown: Julia code is parsed into an AST, which is lowered to Julia IR, for which Julia Codegen produces LLVM IR, which is further compiled by LLVM Codegen. On the bottom, the figure zooms in on additional stages that are executed on the Julia IR. These are macro expansion and control flow destructuring to yield untyped IR, type inference and constant propagation that produce typed IR, inlining to obtain inlined IR and finally \changed{scalar replacement of aggregates (SROA)} and simplification to yield optimized IR.}
    \label{fig:julia_compilation}
\end{figure}

\changed{Fig.~\ref{fig:julia_compilation} shows Julia's just-ahead-of-time compiler flow. At run time, when a method needs to be dispatched and executed, this flow is invoked on it. We will call this the current method. The compiler first lowers the method's AST into an SSA-based IR. Because all operators are functions, that IR code consists mainly of control flow, phi-nodes, literals, assignments to SSA values, and function calls. It hence directly reflects all expressions from the source code, including all called functions and operators, the main difference being that each sub-expression result is assigned to its own SSA value in the IR.
On this IR, the compiler applies type inference based on the run-time argument types, constant propagation, method inlining, and some other basic optimizations such as scalar replacement of aggregates (SROA). From this Julia IR, the compiler then generates LLVM IR, to which many more optimizations are applied by the LLVM compiler before it generates assembly code for the target architecture.}

\changed{
\paragraph{Cranelift and Acyclic E-Graphs} Cranelift~\cite{fallinAegraphsAcyclicEgraphs2023}, a compiler back-end mostly used to generate code from Rust and WebAssembly, is the first production compiler to use e-graphs for code optimization.
As Cranelift aims at JIT compilation, where compilation time is critical, they have introduced \emph{acyclic e-graphs} (ægraphs). Like regular e-graphs, these represent multiple equivalent expressions in e-classes.
But, while regular EqSat fully propagates the effects of introduced equalities, this is not the case with ægraphs.
Equalities are greedily applied, and e-graph invariants are not fully restored.
This means that some rewrites are potentially not discovered, but the compilation time is shorter and more predictable.}

\changed{More relevant to our work is the introduction of the \emph{CFG skeleton}, a data structure kept alongside the e-graph to keep track of control flow instructions and instructions with side effects, two constructs that do not fit the EqSat paradigm for optimizing expressions.
Its workings and how we leverage this technique in our work will be explained in Section~\ref{sec:conversion}.
}
\changed{Because e-graphs in Cranelift are used in the compiler back-end, they model expressions extracted from the IR after inlining, which can be relatively large, but mainly contain simple, low-level instructions, many of which are pure. Cranelifts' e-graphs and its use of the CFG skeleton are hence designed to conserve all control flow and side effects, instead focusing on the rewriting of pure expressions in between control flow and operations with side effects.}

\section{A Rewrite-Based Compiler Middle-End}
\label{sec:rewriter}
Our system allows programmers to specify custom rewrite rules \changed{in Julia syntax}.
These rules are then applied on an e-graph representing the code of a method. Once the rules are applied and the e-graph is saturated \changed{or a timeout has been reached}, the compiler will try to extract the optimal code embedded within that e-graph. 

\changed{Rules have a left-hand side and a right-hand side. Each side consists of a regular Julia expression, i.e., of a tree of function calls, plus variables and/or literals.} Two example rules are 
$
\code{sin(\char`\~x::Number)\char`\^2 + cos(\char`\~x::Number)\char`\^2} \rightarrow \code{1}
$ and 
$\code{translate(\char`\~p::Vec2D, 0, 0)} \rightarrow \code{\char`\~p}
$.
In these rules \code{sin}, \code{cos}, \changed{\code{\^}, \code{+},} and \code{translate} are the names of the called functions. The \code{\char`\~} prefix identifies pattern variables in a rule. These can match any input values of the expression's function calls, be it the values of other expressions, variables, or literals. The \code{::} notation is used to specify the types of values for which the rule can be applied. If no type is specified, values of any type can be matched. 
A rewrite rule can be applied whenever the pattern of function calls, literals, and variables on its left-hand side matches an IR code fragment. 

\changed{As expressions in Julia source code and the SSA IR produced for them contain the same function calls, it is straightforward to search for such matches in a method's SSA IR code. In other words, there is no need to have rule authors express equivalences in terms of IR constructs. They can instead express them in regular source code expressions similarly to how those would occur in source code to be optimized.}
This syntax using regular Julia code expressions, with the addition of unbound, optionally typed variables, covers all rewrites where a single expression tree is replaced with a new expression tree.
This syntax does not allow rules matching \textit{multi-patterns}~\cite{jiaTASOOptimizingDeep2019,MLSYS2021_cc427d93}. Such patterns can match multiple expression trees and rewrite them each, potentially reusing variables in multiple output expressions.

\changed{As in many existing compiler middle-ends, the optimization based on these rewrite rules/patterns will operate on an intermediate representation (IR) that supports optimizations while still preserving high-level language-specific information~\cite{lattner2020mlir,liMirCheckerDetectingBugs2021,swiftSIL}. In our case, this is the Julia compiler's SSA-based IR~\cite{bezansonJuliaFastDynamic2012}.

A consideration to be made for pattern rewriting is at what point in the compilation to apply patterns.
Concretely, patterns can be matched on IR before or after function calls have been inlined.
Before inlining, the current method's IR still contains function calls as they were written by the programmer. At that stage in the compilation, high-level rewrite rules can typically be matched and applied, i.e., rules involving functions of high-level, abstract, and domain-specific APIs.
However, some rewrite opportunities could then remain hidden behind call barriers. This can happen, e.g., when APIs with domain-specific function names wrap more generic computations such as underlying linear algebra computations. If the programmer used the domain-specific wrapper APIs while the rewrite patterns are expressed in terms of the underlying linear algebra operations, those patterns will not be found in the IR code before inlining.

Ideally, when exploring rewrite opportunities, the compiler should hence consider all possible combinations of inlined and not-yet-inlined functions. Support for such an exploration is for future work. Instead, our current prototype implementation performs rewriting optimization only before inlining. We note that many interesting rewrites that are not covered by classical compiler optimization occur before inlining, when high-level API usage has not yet vanished from the code. For this reason also, the example snippets of IR in this paper contain code that has not been inlined yet.
}

\subsection{From IR to E-Graph and Back Again}
\label{sec:conversion}
\begin{figure}[t]
    \centering
\includegraphics[width=\linewidth]{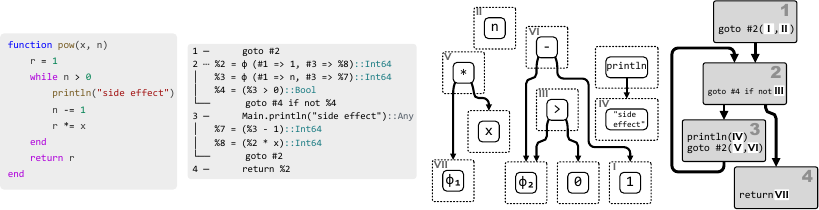}
    \caption{From left to right: a Julia method defined for the function \code{pow} that computes a number's power, the corresponding Julia IR that is generated for two integer arguments, and the e-graph and CFG skeleton for this method.}
    \Description{Top left: a simple Julia method for function pow that computes x to the power n and that prints a message to illustrate the impact of side effects. Top right: the corresponding Julia IR that is generated if the pow method is generated for two integer arguments. It consists of four basic blocks, one of which includes two phi-functions. Bottom left: the e-graph for the expressions in the method consisting of 11 e-classes, including classes for the phi-functions and for the side effect. Bottom right: the CFG skeleton of the method, which looks like the CFG itself, and in which each node refers to certain e-classes in the e-graph where the control flow depends on expressions modeled by those e-classes.}
    \label{fig:conversion}
\end{figure}
To apply rewrites on the Julia IR of a method, we first convert it to an e-graph and a CFG skeleton.
For this, we leverage methods first introduced in the Cranelift compiler~\cite{fallinAegraphsAcyclicEgraphs2023}. 
\changed{Fig.~\ref{fig:conversion} shows an example Julia function \code{pow} together with its SSA IR, e-graph, and CFG skeleton.
The structured control flow present in the original source code has been lowered into four basic blocks in the Julia IR, each terminated by an instruction that jumps to another block or returns.

Each e-node in the e-graph corresponds to an SSA value in the IR. For example, the value \code{\%7} in the IR corresponds to the e-node labeled \code{-}.
The dotted boxes around the e-nodes represent the different e-classes.
Since no equalities have been registered yet in this e-graph, each e-class consists of exactly one e-node.
In this representation, the phi-nodes are leaf nodes, i.e., their dependencies are not included in the e-graph. This is because an SSA phi function introduces a branch in an expression that cannot be represented in an e-graph directly. The lack of phi-node information implies (i) that rewrites including \code{phi}-node dependencies are not supported, and (ii) that the e-graph by itself does not contain sufficient information to reconstruct all control flow of the original code.
For (i), an example of a useful rewrite that is not supported is to reduce a phi-node to a value if all possible values in the phi-node are equivalent.

For (ii), it is the CFG skeleton that provides the extra information needed to reconstruct valid IR. 
To that extent, the CFG skeleton contains all operations responsible for control flow and all operations with side effects, including function calls that have side effects. All operations in the skeleton reference the e-classes they depend on. To visualize these references, we use the roman numerals in Fig.~\ref{fig:conversion}. Each roman numeral labels an e-class of which the corresponding SSA value is used in an operation with side effects, or in a control flow statement. For example, since a call to \code{println} has side effects, it is stored in the third block of the CFG skeleton, referencing e-class {\bfseries IV}.
}

\changed{Cranelift's SSA IR format, like MLIR~\cite{MLIRRationale} and Swift~\cite{swiftSIL}, uses basic block arguments.
These offer an alternative to phi-nodes for representing control-dependent data flow in SSA IRs. Whereas phi-nodes represent such data flow within a basic block, basic block arguments represent such data flow at the basic block level.
Even though Julia IR contains phi-nodes, we opt to use block arguments in the CFG skeleton because all information regarding out-of-block uses of expressions that need to be materialized within a block is then explicitly represented within that block itself.

To illustrate, the references to e-classes {\bfseries I}, {\bfseries II}, {\bfseries V}, and {\bfseries VI} in the two \code{goto \#2} statements in the skeleton are basic block arguments.
They are passed as arguments to control flow statements and can be references in the destination block, similarly to how functions have arguments.
The relation between basic block arguments and phi-nodes in the original Julia IR is simple.
For example, the two phi-nodes in block \code{\#2} in the Julia IR correspond to the two basic block arguments passed in the CFG skeleton by blocks \code{\#1} and \code{\#3}.
Note that while the e-graph in Fig.~\ref{fig:conversion} contains nodes labeled phi, these are purely labels and do not carry the information contained in SSA phi-functions. The phi-nodes in the e-graph can be interpreted as being the corresponding basic block arguments. 
}

It is the CFG skeleton that allows the conversion back to valid Julia IR. As long as each statement in the CFG skeleton is \textit{elaborated}, meaning that that statement and all of its dependencies are materialized in the generated IR, the resulting program has the same control flow and side effects as the original program.

The need to support control flow stems from a number of common code patterns of expressions that are split over multiple basic blocks for the sake of code conciseness, readability, and maintainability. The most important case concerns if-then-else patterns in which different expressions are computed under different conditions, and of which a common subexpression is computed beforehand. Five idioms for which such patterns occur are the following:
\begin{description}
\item [Keyword Arguments (kwargs)] It is quite common that part of a method depends on a kwarg, e.g., to perform extra checks or to modify the behavior. See, e.g., the use of the \code{flipkernel} kwarg in NNLib.jl~\cite{NNlib}.
\item [LAPACK-style wrappers with \code{uplo}/\code{fmt}/\code{trans}/... chars] BLAS/LAPACK-like libraries often use characters to indicate the format of an array (transposed or not, upper/lower triangular, etc.). Such checks are then independent of the values calculated beforehand. For example, in the LinearAlgebra.jl standard library~\cite{LinAlg.jl}, the costly operation \code{blks = findall(...)} is first computed, after which there is a branch based on \code{uplo}, followed by a reuse of \code{blks}. Another example is CUDA.jl's CUSPARSE wrappers: Plenty of API calls (which are a prime use case for e-graph-based rewriting) occur outside of \code{fmt} checks, others occur inside such checks~\cite{CUDA1.jl}.
\item [API Version Checks] Continuing the subject of API calls, it is common that certain API uses are conditional on the version of the library. For example, CUDA.jl includes checks such as \code{CUSPARSE.version()} $\geq$ \code{v"11.7.2"}~\cite{CUDA2.jl}.
\item [Low-level Math] In scalar ‘mathy’ code, this pattern is also common. The absolute cost of operations outside the if is then often not that great, but when it comes to optimizing very high-performance code, it can make a difference. Some examples can be found in the standard float.jl~\cite{float.jl} and math.jl~\cite{math.jl} libraries.
\item [Imperative Type Checks] For readability or to avoid too many methods, imperative type checks are also sometimes used in Julia instead of multiple dispatch. Several examples can be found in the standard libraries~\cite{CUDA3.jl,abstractarray.jl,intfuncs.jl}.
\end{description}

All of these patterns involve divergent control flow, which our work can handle without problems.

To convert an e-graph back to IR, the statements in the CFG skeleton are then used as starting points. This ensures that side effects and control flow are kept intact.
IR is created by materializing each statement in the CFG skeleton as well as all its dependencies in the e-graph.
Using the \textit{scoped elaboration} algorithm~\cite{fallinAegraphsAcyclicEgraphs2023}, materialization occurs only for values that have not yet been materialized in statements that dominate the current statement in the CFG. Otherwise, the previously materialized SSA value is reused.

Our implementation uses Metatheory.jl~\cite{cheliMetatheoryjlFastElegant2021}, a Julia EqSat package based on egg~\cite{willseyEggFastExtensible2021}.
\changed{At its core, it provides an efficient implementation of the e-graph datastructure and EqSat algorithms, and a system that allows to register new types of e-nodes.
In our work, we use this system to implement e-nodes that represent Julia IR statements.
Metatheory.jl also offers macros to easily specify rewrite rules with a high-level, declarative syntax.
We extended this rule syntax to support type annotations, as will be discussed in Section~\ref{sec:types}.
Lastly, Metatheory.jl contains a generic extraction algorithm. However, this algorithm is unsuited for our problem, as extraction needs to take into account the information from the CFG skeleton as well. In Section~\ref{sec:extraction} we will discuss our extraction approach.
}

\subsection{Rules and Types}
\label{sec:types}
\changed{With multiple dispatch,} the semantics of a function call are decided not only by syntax (i.e., the function name), but also by the types and counts of their arguments. This necessitates a way to represent this information in rewrite rules.
For example, a user might want to rewrite matrix multiplications into a function call to some external BLAS library.
This means that a call to \code{Base.:*}, a base library function, should be rewritten, but only if its arguments are matrices.
Users can specify type constraints in the rewrite rules by associating a type with an unbound variable.
For example, a rule for rewriting a matrix multiplication can be written as
$
\code{\char`\~A::Matrix * \char`\~B::Matrix} \rightarrow \code{blas\_call(\char`\~A, \char`\~B)}.
$

\changed{E-class analyses~\cite{willseyEggFastExtensible2021} \changed{offer a framework} to associate and propagate lattice information in an e-graph. 
}
We designed a novel e-class analysis for tracking types in the e-graph.
For each e-class, this analysis maintains the most specific type that is compatible with all the e-nodes in that class.
During e-graph construction, this type is simply the same as the types of statements in the Julia IR.
When two e-classes with a different associated type are merged, the resulting e-class is determined by the \code{typejoin} function.
For rewrite rules that introduce a new function call, such as the rewrite of a matrix multiplication into \code{blas\_call}, we run Julia's type inference using the types stored in the e-classes.
Similarly, if the type associated with an e-class changes, we rerun type inference for all e-classes containing e-nodes that depend on the changed e-class.
\changed{
As an optimization, for a small amount of different types that are joined, we construct a \code{Union} type that keeps track of all individual possible types. When the union grows too large, we fall back on a single type that is their most specific joint supertype.
For example, if a value of type \code{Integer} is introduced in an e-class with associated type \code{Float64}, the new associated type becomes \code{Union\{Integer, Float64\}}. If, throughout EqSat, this union grows further, the associated type could become \code{Number}.
Tracking small unions of specific types instead of joining them immediately into their joint supertype enables additional rewrite rules bound to the more specific types.}

\changed{Rewrite rules that change the type of a value produced by an expression can cause problems. If the value is used by a function that has no method defined for the new type, no valid code can be produced anymore. Our current implementation detects this when the type information of parent nodes is recomputed. When type inference finds no method for arguments with the new type, an error is thrown.
It is the responsibility of the rule author to either ensure that rules only introduce new types that are compatible with all the functions that depend on the value, or that new methods are defined to make it so. In future work, we will implement the necessary support to skip the application of rewrite rules that can result in the production of invalid code.}

\subsection{Constants in the E-Graph}
Rewrite rules are usually symbolic in nature.
They match an expression and replace that expression with a new one, plugging in any matched variable.
Modern EqSat frameworks such as egg~\cite{willseyEggFastExtensible2021} and Metatheory.jl~\cite{cheliMetatheoryjlFastElegant2021}, which is used in this work, also support \textit{dynamic rewrites}. When dynamic rewrite rules are applied, the \changed{expression in the} right-hand side is first \changed{evaluated} (potentially using additional e-class analysis data) and the original code is rewritten into the result of that execution, rather than into the literal right-hand side expression itself.

When literals are part of the e-graph, dynamic rewrites can be used to perform simple constant \changed{folding}.
In normal Julia code, literals of primitive types and composite types with known size can appear directly in the IR.
The values of these literals are embedded in the e-graph.
The e-classes containing the corresponding e-nodes of the literals have an inferred type associated with them, but compared to e-classes containing nonliteral e-nodes, the actual value of that type is available as well. To support constant folding of literals, we extend rewrite rule type annotations with a special \code{Comptime\{T\}} type that \changed{only allows to match against literals of type \code{T}.} Using $\Rightarrow$ instead of $\rightarrow$ to signify dynamic rewrite rules, a rule such as 
{
$  
\code{\char`\~a::Comptime\{Integer\} + \char`\~b::Comptime\{Integer\}} \Rightarrow \code{\char`\~a + \char`\~b}
$ 
}
will match any addition between two integer literals and insert the \changed{constant-folded} result of this addition in the respective e-class. 

Dynamic rewrite rules also open opportunities for partial evaluation-like optimization~\cite{10.5555/286076,futamuraPartialEvaluationComputation1999} beyond the constant propagation capabilities of the Julia compiler. 
The compiler does not propagate dynamic array arguments as their values could be mutated at any time.
With dynamic rewrite rules, a programmer can override this default behavior.
To achieve this, we extended the rewrite rule format with rules of the form $ \code{function.arg[2]}\Rightarrow \code{W}.$
This rule introduces an equivalence between the function's second argument in its e-graph and the constant \code{W}. A constant array holding \code{W}'s contents is added as an equivalent e-node to that argument's e-class in the e-graph, to which dynamic rewrite rules can then be applied. An interesting case is, for example, when \code{W} is a matrix holding the constant weights of a neural network layer.
Computations that only depend on \code{W} or other compile-time-known variables can then be partially evaluated.

\subsection{Rewrites in the Presence of Side Effects}
\changed{As discussed in Section~\ref{sec:conversion}, the CFG skeleton is used to ensure that all statements with side effects, such as the \code{println} in the example of Fig.~\ref{fig:conversion}, are materialized when the final optimized program is extracted.}

We \changed{leverage} Julia's built-in static effect analysis to determine whether a function call has side effects. This analysis determines several properties for each IR statement.
The program property \code{effect\_free} signals whether a statement is ``free from externally semantically visible side effects''~\cite{juliaEffectDocs}.
\changed{For our purpose, however, this analysis is often too strict and too imprecise, resulting in some functions not being labeled \code{effect\_free}, even though they do not have side effects that require them to remain in a program.
\changed{For example, many functions that allocate memory are not marked effect-free, even if the allocated memory never gets accessed again}.
As a result, even if rewrite rules subsequently introduce equalities that would make calls to these functions superfluous, they will still be included in the optimized code.
In short, there is a mismatch between Julia's effect analysis and the desired freedom to operate of our rewrite system.}

In Cranelift, side effects are less of a concern because the code it rewrites is lower level and typically contains a large amount of simple, low-level instructions, many of which are pure.  In high-level Julia code, by contrast, the IR we work on generally is much smaller and consists only of a few calls to high-level functions/methods.
Because these functions/methods have more complex behavior than simple, low-level instructions, they also often include behavior with potential side effects according to Julia's built-in analysis. 

\changed{To overcome this mismatch when rewriting Julia code, we allow users to prefix rewrite rules with the \code{relaxed} keyword.
Whenever a relaxed rule matches code, all the function calls that are part of the CFG skeleton and that occur in the left-hand side of the rule are ``detached'' from the CFG skeleton.
When the final optimized code is extracted from the e-graph, detached statements are not forced to be materialized. We call this technique \emph{CFG skeleton relaxation}.}

\changed{Technically, it is possible to write relaxed rules that get rid of side effects that are depended upon by other parts of the code, thereby breaking its intended semantics. We note, however, that all practical rewrites that we encountered and that needed relaxation as we will discuss in Section~\ref{sec:usecase}, did not introduce problems.

Finally, we note that semantic rule verification is outside the scope of this work. As a rewrite rule might be correct in one domain but incorrect in another, such as reduced precision in deep learning, correctness is up to the rule author.}

\subsection{Extraction}
\label{sec:extraction}
\begin{figure}[t]
    \centering
\includegraphics[width=0.85\textwidth]{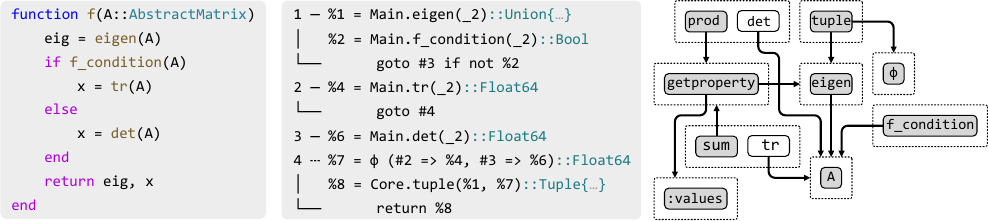}
    \caption{Example method for function 
    \code{f} for which EqSat reveals opportunities for reuse. \captioncode{tr} and \captioncode{det} can be rewritten as a sum and a product over the eigenvalues stored in \captioncode{eig}. Optimal e-nodes for extraction are shaded gray.}
    \Description{Example function where EqSat reveals opportunities for reuse. Based on some condition, the function returns either the trace of matrix A or its determinant. The invocations to the functions tr and det that compute the trace respectively the determinant, can be rewritten as a sum and a product over the eigenvalues stored in eig, which is the variable in which the eigenvalues are already stored after a call to the eigen function. The Julia IR is shown next to the source code, and the e-graph is chosen, which consists of 9 e-classes totaling 11 e-nodes. The nine optimal e-nodes, one for each e-class, for extraction are shaded gray.}
    \label{fig:extraction}
\end{figure}
When Julia IR is generated from an optimized e-graph and its corresponding CFG skeleton, each statement in the CFG skeleton is considered a root node for e-graph extraction. This means that, in contrast to \changed{vanilla egraph extraction}, we essentially need to run multiple extractions \changed{on the single e-graph} to recover our program~\changed{\cite{cowardAutomatingConstraintAwareDatapath2023}}.
These extractions are not independent of each other.
Take, for example, the code and corresponding IR and e-graph in Fig.~\ref{fig:extraction}.
The method \code{f} computes and returns two different values \code{eig} and \code{x} derived from matrix \code{A}. \code{eig} stores the eigen vectors and values, and \code{x} stores the trace or the determinant of the matrix, depending on control flow.
In this example, two rules from a collection of linear algebra identities have been applied: 
$\code{tr(\char`\~A::AbstractMatrix)} \rightarrow \code{sum(eigen(\char`\~A).values)}$ and  $\code{det(\char`\~A::AbstractMatrix)} \rightarrow \code{prod(eigen(\char`\~A).values)}$.
These give rise to the extra e-nodes in the e-classes containing \code{det} and \code{tr}.
Intuitively, we can see that an optimal extraction might be one in which the result of the call to \code{eigen} is reused to compute the determinant and/or trace instead of explicitly calling those functions.
Indeed, that extraction is valid since the call to \code{eigen} dominates both the call to \code{det} and to \code{tr}, so its result is available at these call sites.

We formulate the problem of finding an optimal extraction as an ILP problem.
\citeauthor{heImprovingTermExtraction2023}~\cite{heImprovingTermExtraction2023} propose an ILP description for e-graph extraction that ensures an acyclic extracted graph, which is important for further elaboration to produce linearized IR.
We adapt and extend their description to take into account e-node reuse from dominating extractions.

\begin{figure}
    \centering
    \includegraphics[width=6cm]{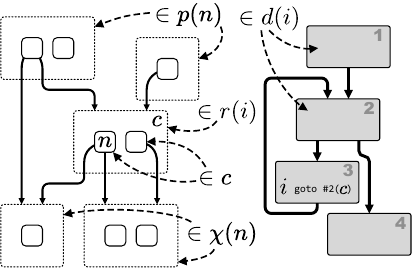}
        \captionof{figure}{\emph{Left:} e-graph with 5 e-classes. E-node $n$ has 2 child e-classes. Both e-nodes in $c$ have the same two e-node parents. \changed{\emph{Right:} statement $i$ in block 3 of the CFG skeleton is dominated by all statements in dominating blocks, and depends on e-class $c$ in the e-graph.}}
        \label{fig:egraph_notation}
\end{figure}

Let the e-graph be represented by a set of e-nodes $N$, and a set of e-classes $C$. Let $\left\{n | n \in c\right\}$ be the set of e-nodes contained in e-class $c \in C$, and let $\chi(n)$ be the set of children e-classes of a particular e-node $n \in N$.
\changed{We associate with each e-class $c$ a set of e-nodes $p(c)$ defined as $ \{n \mid c \in \chi(n)\}$.
That is, the set of \emph{parent} e-nodes $n$ that have a child e-class $c$.
}
Fig.~\ref{fig:egraph_notation} shows an e-graph \changed{and CFG skeleton} in which the different parts have been annotated with this notation.

We represent the statements contained in the CFG skeleton as the set $I$.
All of these statements cause a different expression to be extracted from the e-graph, rooted at each $i \in I$.
For each statement $i \in I$, the statements that dominate $i$ are represented by the set $d(i)$.
Each statement $i$ depends on a number of e-class arguments represented by the set $r(i)$.

Finally, we associate a cost $M(n)$ with every e-node $n$ such that our optimization problem becomes\footnote{Changes to the formulation by \citeauthor{heImprovingTermExtraction2023}~\cite{heImprovingTermExtraction2023} are highlighted in blue.}
\setlength{\columnsep}{15pt}
\begin{multicols}{2}
\begin{flalign}
&\quad \min \sum_{\highlight{i \in I}, n \in N} M(n) \cdot w_n^{\highlight{(i)}} \quad \underset{\highlight{\forall i \in I}}{\text{subject to}}&\nonumber\\
&\sum_{n' \in c}\left( w^{\highlight{(i)}}_{n'} + \highlight{\sum_{j \in d(i)} w_{n'}^{(j)}}  \right) \ge w_n^{\highlight{(i)}}, \forall \begin{array}[t]{l} n \in N, \\ c \in \chi(n) \end{array}\label{eq:constraint1}&\\
&\highlight{w^{(i)}_{n} \le \sum_{n' \in p(\changed{c})} w^{(i)}_{n'}, \quad \forall \changed{c \in C-r(i), n \in c}}&\label{eq:constraint2}\\
&\sum_{n \in c} \left( w_n^{\highlight{(i)}} + \highlight{\sum_{j\in d(i)}{w_n^{(j)}}} \right) \geq 1, \quad \forall c \in r\highlight{(i)}&\label{eq:constraint3}
\end{flalign}

\columnbreak

\hfill

\begin{flalign}
&\left(1 - v_{\Psi,c}^{\highlight{(i)}}\right) + \left(1 - w_n^{\highlight{(i)}}\right) \geq 1, \forall  
\begin{array}[t]{l}
\Psi \in \mathcal{A},  
c \in \Psi,\\ 
\quad \quad \quad n \in \mathcal{N}_{\Psi}^{c} 
\end{array}\\
&v_{\Psi,c}^{\highlight{(i)}} + \sum_{n \in \mathcal{N}_{\Psi}} w_n^{\highlight{(i)}} \geq 1, \quad \forall \Psi \in \mathcal{A}, c \in \Psi\\
&\sum_{c \in \Psi} v_{\Psi,c}^{\highlight{(i)}} \geq 1, \quad \forall \Psi \in \mathcal{A}\\
&w_n^{\highlight{(i)}}, v_{\Psi,c}^{\highlight{(i)}} \in \{0, 1\}, \quad \forall \Psi \in \mathcal{A}, c \in \Psi, n \in N
\end{flalign}
\end{multicols}

Where, for each e-node $n$ and statement $i$ from the CFG skeleton, we have introduced a binary variable $w_n^{(i)}$.
These variables indicate whether the e-node is selected for the extraction rooted at statement $i$.
Furthermore, following the work by \citeauthor{heImprovingTermExtraction2023}~\cite{heImprovingTermExtraction2023}, each $\Psi \in \mathcal{A}$ is a \textit{class cycle}, a collection of classes that contain e-nodes that form a cycle by depending on other classes in the cycle.
The e-nodes in an e-class $c$ that participate in a class cycle $\Psi$ are represented by the set $\mathcal{N}_{\Psi}^{c}$.
For each class $c$ in each class cycle $\Psi$, we introduce binary variables $v_{\Psi,c}^{(i)}$ to enforce acyclicity in the extracted graph.
$v_{\Psi,c}^{(i)}$ is assigned 1 if none of the e-nodes in e-class $c$ that participate in the class cycle are selected.
For a more complete overview and proof of acyclic extraction, we refer to the work by \citeauthor{heImprovingTermExtraction2023}~\cite{heImprovingTermExtraction2023}.
To take into account the dominance information from the CFG skeleton, we adapted their formulation in a few ways.

We adapt constraint (\ref{eq:constraint1}) to ensure that an e-node can only be selected if for each child e-class an e-node is also selected, \emph{or if an e-node is already selected in a dominating extraction $j$}.
To prevent e-nodes from being selected in dominating extractions when they are not needed there, we introduce constraint (\ref{eq:constraint2}).
This constraint prevents an e-node from being selected if there is no e-node selected at the same extraction that depends on that e-node.
Constraint (\ref{eq:constraint3}) ensures that at least one e-node in each e-class that is referenced by a root in the CFG skeleton is truly selected for extraction.
Compared to the original ILP formulation by \citeauthor{heImprovingTermExtraction2023}~\cite{heImprovingTermExtraction2023}, we added an additional term to account for the fact that an e-node might already have been selected in a dominating extraction.

The other constraints are completely analogous to the constraints by \citeauthor{heImprovingTermExtraction2023}~\cite{heImprovingTermExtraction2023} except for the addition that each constraint is added separately for each possible extraction root $i \in I$.

We implemented this ILP formulation in JuMP~\cite{lubinJuMP10Recent2023}, a domain-specific language embedded in Julia to express and solve mathematical optimization problems.
For a solver, we resorted to HiGHS~\cite{huangfuParallelizingDualRevised2018}.
By default, our cost function $M(n)$ currently defaults to 1, reducing the optimization problem to choosing the least number of e-nodes.
We allow users to manually alter the cost for calls to particular functions to overwrite this behavior.


\section{Use Case Demonstration}
\label{sec:usecase}
\begin{figure}
    \centering
    \includegraphics[width=6.8cm]{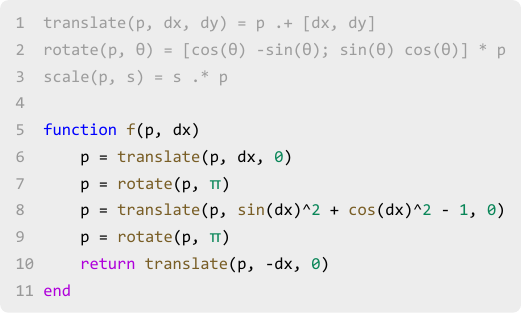}\caption{Domain-specific code applying 2D transformations. }
    \Description{Example of domain-specific code used to apply 2D transformations to a point.
    The method performs a series of transformations: a horizontal translation over distance dx, a rotation over pi radians, a translation over cosine squared plus sine squared minus one, which equals zero, another rotation over pi radians, and a translation over minus dx. }
    \label{fig:2D_transformations}
\end{figure}

To demonstrate the capabilities of our work, this section presents a number of use cases, focusing on our main goal of allowing users and domain experts to express and exploit optimization opportunities in a convenient, flexible way. In other words, our use cases aim to demonstrate that our work can give developers the necessary control over optimizations without having to become compiler experts and without restricting the flexibility and productivity they have come to expect from the Julia ecosystem. In contrast, our work aims to give developers more flexibility, meaning less restrictions, to write code without incurring a reduction in optimization opportunities.  


\subsection{Domain Specific Rewrites}
\label{sec:2D_transformations}
Fig.~\ref{fig:2D_transformations} shows a method that applies a sequence of similarity transformations on a 2D point.
On close inspection\changed{, under the assumption that exact floating point results must not be maintained}, application of geometric properties, trigonometric identities, and algebraic simplifications can lead to the reduction of all these transformations to a single identity transform.
Indeed, the translation on line~8 can be simplified to \code{p} by applying the following three rules:
$\code{sin(\char`\~x::Number)\char`\^2 + cos(\char`\~x)\char`\^2} \rightarrow \code{1}$, 
$\code{\char`\~x::Number - \char`\~x} \rightarrow \code{0}$, and 
$\code{translate(\char`\~p::Vector, 0, 0)} \rightarrow \code{\char`\~p}$. 
After this translation has been eliminated, the two rotations by $\pi$ radians similarly end up not transforming the point.
Lastly, the first and last rotations now cancel each other out as well.
After all the corresponding rewrites, the optimized method simply returns its argument \code{p}.

This example also illustrates the necessity of CFG skeleton relaxation.
The Julia compiler's effect inference fails to reason through the rotation matrix construction in the \code{rotate} function on line~2 in Fig.~\ref{fig:2D_transformations}.
\changed{As a result, the allocation of the rotation matrix is conservatively considered to have side effects, which means that, by default, even if an optimization makes the rotation matrix obsolete, the matrix will still be materialized by the optimized code.}
The rewrite rule author is hence required to allow CFG skeleton relaxation for rules with \code{rotate} by prefixing those rules with \code{relaxed}.

\subsection{Aiding Multiple Dispatch}
Instead of using rewrites purely for simplifying code, they can also be used to generate code that provides additional information to the Julia compiler' type system to trigger the execution of better performing methods through multiple dispatch. 
An example in the domain of linear algebra is to automatically wrap certain matrix expressions in a new, more specific type that allows more efficient implementations for subsequent computations.
Take, for example, the expression $A + B^T$ between two regular matrices $A$ and $B$.
For the case where $A$ and $B$ are the same matrix, that is, $A+A^T$, we know that the result is a symmetric matrix.
We can encode this fact by applying the following rewrite rule:
$ 
\code{\char`\~A::AbstractMatrix + transp(\char`\~A)} \rightarrow\code{Symmetric(\char`\~A + transp(\char`\~A))}
$.

The \code{Symmetric} function from the \code{LinearAlgebra.jl} package takes the upper triangle of its argument and uses that to efficiently represent a symmetric matrix. 
More importantly, it returns a value of type \code{Symmetric}. Other functions in that linear algebra package, e.g., for solving eigenproblems, have specialized methods for symmetric matrices. After rewriting with the above rule, those specialized methods will now be called automatically thanks to the multiple dispatch system knowing that they are invoked on the type \code{Symmetric} instead of on a more generic matrix type. 
Another example is the expression $P^T B P$, which is symmetric if B is symmetric as well, as can be encoded with the rule
$ 
\code{transp(\char`\~P::AbstractMatrix) * \char`\~B::Symmetric * \char`\~P} \rightarrow\code{Symmetric(transp(\char`\~P) * \char`\~B * \char`\~P)}
$.

Both examples build on the abstractions of the Julia standard library, but the same concepts can also be applied to operations and types in other, external packages.

\subsection{Multi-Line Broadcast Fusion}
\label{sec:multi-line broadcast fusion}
\begin{figure}
\begin{minipage}[t]{0.47\textwidth}
\centering
    \includegraphics[width=5.5cm]{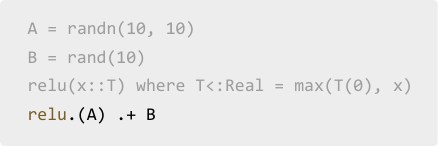}
    \caption{Broadcasting syntax: the \captioncode{relu} function is broadcasted over \captioncode{A} and \captioncode{B} is added to the result using broadcasting addition.}
    \Description{Julia code that uses broadcasting syntax: using the dot operator, an elementwise ReLu operation and an elementwise addition are applied to vectors A and B. In other words, the relu function is broadcasted over matrix A and vector B is added to the result using broadcasting addition.}
    \label{fig:broadcast_dot}
\end{minipage}\quad\quad\begin{minipage}{0.47\textwidth}
    \centering
    \includegraphics[width=6.5cm]{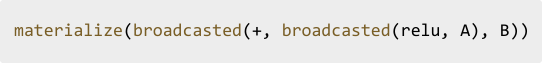}
    \caption{Syntactic lowering applied for the broadcasting expression in Fig.~\ref{fig:broadcast_dot}. \captioncode{broadcasted} builds a lazy representation of the broadcast operation. \captioncode{materialize} allocates the result array and computes its values.}
    \Description{Julia code example of materialize(broadcasted(+,broadcasted(relu,A),B)) to illustrate syntactic lowering applied for the broadcasting expression in the previous figure showing a broadcasting expression. The broadcasted function builds a lazy representation of the broadcast operation. The materialize function allocates the result array and computes its values.}
    \label{fig:broadcast_expanded}
\end{minipage}
\end{figure}

Julia's broadcasting mechanism allows users to apply functions element-by-element on one or more arguments containing multiple elements.
Syntactically, this is done by adding a dot (\code{.}) between the function and its argument list.
The top of Fig.~\ref{fig:broadcast_dot} shows two broadcast operations.
\code{relu.(A)} applies the scalar \code{relu} function to each element of the matrix \code{A}.
Next, the \code{.+} call computes the addition between the result and a vector \code{B}.
Although the two arguments to \code{.+} have a different shape, the operation can take place because the shapes are compatible.

The Julia compiler transforms expressions containing this special \textit{dot syntax} into calls to the standard library functions \code{broadcasted} and \code{materialize}.
Fig.~\ref{fig:broadcast_expanded} shows the result for the expression in Fig.~\ref{fig:broadcast_dot}.
A call to \code{broadcasted} builds a lazy representation of the broadcast result. Different lazy broadcast objects can be nested; it is only when \code{materialize} is called that the final result object is allocated and filled with computed values.
As such, different operations within multiple \code{broadcasted} calls are fused when \code{materialize} is called.
This leads to less memory being used because temporary arrays are not materialized and potentially better run-time performance because there are fewer function calls.
This operator fusion is especially beneficial for code that is executed on the GPU, as kernel launches can take a significant amount of time, and because kernel fusion can significantly reduce the required number of memory accesses~\cite{besardEffectiveExtensibleProgramming2019}. 

A limitation of Julia's broadcasting lowering is that fusion cannot occur across broadcast expressions on different lines.
This is because replacing the dot syntax with calls to \code{broadcasted} and \code{materialize} is a syntactical transformation that is performed during AST lowering, which operates statement by statement. A potential optimization is hence to eliminate superfluous calls to \code{materialize}.
Rewrite rules make this easy.
Take, for example, the rule
$ \code{broadcasted(\char`\~f, materialize(\char`\~x))} \rightarrow \code{broadcasted(\char`\~f, \char`\~x)}.$
This rule will remove the intermediary materialization for any call of the form \code{f.(x)} where \code{x} is a variable that was defined on a different line with a broadcast expression itself.
In the example in Fig.~\ref{fig:deep learning original}, fusion will now be applied despite the two dot operators occurring in two different statements on two separate lines. This illustrates how the system we propose not only supports domain-specific optimization but also at the same time enables optimization of common code patterns involving only core Julia primitives. 

An important detail is that here, once more, CFG skeleton relaxation is required, since \code{materialize} ends up calling a foreign C function which the compiler cannot deem effect-free, poisoning the remainder of the effect analysis.
In practice, this means that the rule needs to be prefixed with \code{relaxed}.

Using PkgEval.jl, a tool to automatically evaluate the tests of Julia packages, we evaluated how many times the above rule's pattern occurs in real-life code in 3153 different packages.
In total, we found 514 instances where code containing this pattern was executed, spread over 232 different packages.
This pattern occurs, for example, when a broadcasted expression is used in different branches and is factored out by the programmer as a code simplification measure.
Applying the rewrite rule does not undo the deduplication but only removes the superfluous call to \code{materialize}.
Note that the discussed rule only matches expressions where the outer broadcasted function is a unary function.

For other patterns, e.g., an expression where the outer broadcasted function is a binary function, separate rules need to be written.
Table~\ref{tab:rule counts} lists how many times our analysis found different multi-line broadcasting patterns that can all be rewritten to $\code{broadcasted(\char`\~f, \char`\~x)}$ and $\code{broadcasted(\char`\~binop, \char`\~x, \char`\~y)}$.
The last three patterns are different configurations of broadcasting a binary function. Note that the matches for these three patterns are not independent: a match against the last pattern, which will result in two invocations of \code{materialize} being optimized out, implies the two other patterns can be matched and applied as well, each optimizing out only one \code{materialize} invocation. When the final code is extracted in such cases, the expression from the e-class containing the least materialize calls will be chosen.

\begin{table}[t]
    \centering \small
    \begin{tabular}{l|c|c}
        \hline
         \makecell{\textbf{LHS pattern of rewriting rule }} & \makecell{\textbf{Count}}  & \makecell{\textbf{Distinct Packages}} \\
        \hline
        \smallcode{broadcasted(\char`~f, materialize(\char`~x))} & 514 & 232 \\
        \smallcode{broadcasted(\char`~binop, materialize(\char`~x), \char`~y)} & 1232 & 348 \\
        \smallcode{broadcasted(\char`~binop, \char`~x, materialize(\char`~y))} & 1044 & 283 \\
        \smallcode{broadcasted(\char`~binop, materialize(\char`~x), materialize(\char`~y))} & 231 & 93 \\
        \hline
    \end{tabular}
    \caption{Number of matches for different multi-line broadcasting patterns in an analysis of 3153 Julia packages.}
    \label{tab:rule counts}
\end{table}

\subsection{Extending Broadcasting with Domain-Specific Batched Operations}
\label{sec:fused}
\begin{figure}
\begin{minipage}{0.48\textwidth}
    \centering
    \includegraphics[width=6.6cm]{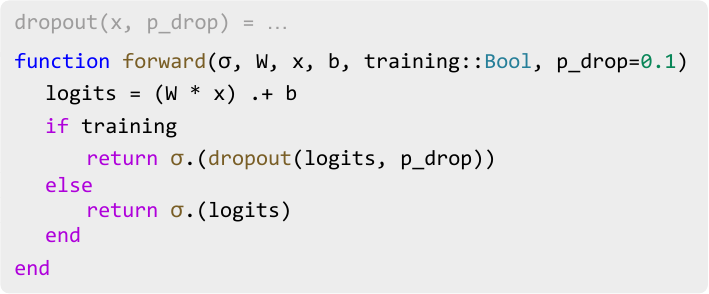}
    \caption{Straightforward code in the core programming language, without needing to use specialized functions with custom APIs.}
    \Description{
    Julia code example of a function called forward that computes the forward pass of a single-layer neural network. Apart from network inputs, weights, and activation function, the forward function takes a boolean argument called training, and an argument called p_drop which is the probability value used in dropout operation.
    The function starts by assigning the logits variable, which is computed by a matrix multiplication between the weights and input, and a broadcasted addition with the bias vector.
    Using an if statement, depending on whether the training flag is set true or false, the dropout function is applied before broadcasting the activation function on the result and returning it.
    }
    \label{fig:deep learning original}
\end{minipage}\quad\quad\begin{minipage}{0.47\textwidth}
    \centering
    \includegraphics[width=7cm]{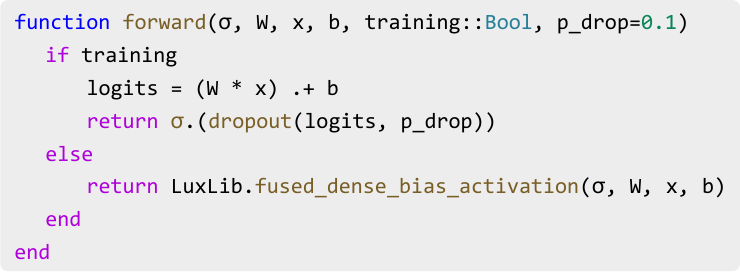}
    \caption{Manually rewritten source code that calls an optimized implementation of certain computations.}
    \Description{
    Julia code of the same neural network computation as the previous figure, after rewrite transformations. Now, the branch where no dropout is applied directly a specialized function called fused_dense_bias_activation. The branch where dropout is applied still first computes the logits variable, applies the dropout function, and returns the value after applying the activation function.
    }
    \label{fig:deep learning rewritten}
\end{minipage}
\end{figure}

In high-performance contexts, programmers often sacrifice code simplicity and readability in pursuit of achieving the highest possible performance.
For example, instead of writing deep learning layers and non-linear activation functions using core language constructs, programmers often resort to using frameworks or libraries that contain optimized implementations of matrix multiplications, activation functions, attention computation, and others~\cite{elshawiDLBenchComprehensiveExperimental2021,parvatSurveyDeeplearningFrameworks2017}.
This works well for most common use cases, but falls apart when a programmer requires custom behavior that is not yet supported by the framework.
In those cases, the programmer is required to rewrite parts of the program to stop using the framework and instead depend on a different framework or the core language.
Instead of having to opt out of framework functionality and rewriting code, one could ideally write simple, readable code and have the frameworks themselves automatically discover opportunities to use their more performant implementations. 
Our system gives framework authors the option for implementing such discovery and optimization almost for free.

As an example, the code in Fig.~\ref{fig:deep learning original} shows an implementation of the forward pass of a single fully connected feedforward layer.
Several existing Julia packages allow us to compute such operations more efficiently~\cite{faingnaertFlexiblePerformantGEMM2022,innesFashionableModellingFlux2018,innesFluxElegantMachine2018}.
For example, the programmer could rewrite the code as shown in Fig.\ref{fig:deep learning rewritten} to use the optimized \code{fused\_dense\_bias\_activation} function from the Lux.jl deep learning framework~\cite{palLuxExplicitParameterization2023}. A disadvantage of such manual rewriting is that the code becomes less readable for someone not familiar with the Lux.jl APIs. Moreover, the use of that specific framework's optimized implementation, which is in essence a decision about a non-functional aspect of the code, is then hard-coded in the code that specifies the functionality. This requires the developers and the maintainers of this code to be knowledgeable in the code's application domain as well as in the Lux.jl framework. Switching to other frameworks in the future will require rewriting code. Clearly, hard coding the dependence on Lux.jl and its APIs in the code has a number of downsides.

With our system, instead of rewriting the original code of Fig.~\ref{fig:deep learning original}, it suffices to add a rewrite rule that maps the original code pattern onto the optimized \code{fused\_dense\_bias\_activation} function: {
\begin{align*}
& \code{materialize(broadcasted(\char`\~$\sigma$, broadcasted(+, \char`\~W::AbstractMatrix * \char`\~x::AbstractMatrix, \char`\~b::AbstractVector)))}
\\
&\quad\rightarrow \code{LuxLib.fused\_dense\_bias\_activation(\char`\~$\sigma$, \char`\~W, \char`\~x, \char`\~b)}.
\end{align*}
}

Combined with the multi-line broadcast rewriting rule explained above, which rewrites the original code into a single line without the intermediary variable \code{logits}, this rule maps the single-line broadcast expression onto a call to the library function \code{LuxLib.fused\_dense\_bias\_activation}, identical to what would be generated for the code in Fig.~\ref{fig:deep learning rewritten}.

This example illustrates how our approach allows for a better separation of concerns. Another advantage is that alternative rewrite rules can be provided for different Julia packages that provide different optimized implementations for similar functionality.
Different rewrite rules also do not obfuscate each other.
All rewrite rules can be represented in the e-graph.
Only during extraction is the final optimized code materialized, optionally based on user-defined extraction costs for different functions.
As such, rewrite rules from different authors are fully interoperable.

Similarly, without changing the source code, different rewrites can still occur by applying different sets of rewrite rules that can, for example, be offered by different Julia packages.

This example also shows how our system supports composing high-level, domain-specific optimizations with optimizations of core Julia primitives. In this way, its scope is much broader than recent EqSat optimizers like Cranelift.

\subsection{Keeping up with evolving and emerging libraries for GEMMs}
\label{sec:gemm}
\changed{As an example of multiple libraries providing complementary functionality, consider General Matrix Multiplications (GEMMs), a cornerstone in many scientific domains. GEMMs are also often combined with element-wise operations or \emph{elops} on their operands or result, for example with non-linear activation functions such as ReLU.

Many libraries exist to perform different GEMM variants~\cite{vanzee2015BLIS,faingnaertFlexiblePerformantGEMM2022,torres2024evaluationcomputationalenergyperformance}, each offering different trade-offs and interfaces. For targeting NVIDIA GPUs, e.g., the cuBLAS library provides the highest performance for basic GEMM expressions on matrices with simple datatypes. It also offers some kernels that fuse often-used elops with the GEMM.
For elops not supported by cuBLAS, the alternative GemmKernels.jl can be used~\cite{faingnaertFlexiblePerformantGEMM2022}. This package leverages Julia's multiple dispatch and GPU programming support~\cite{besardEffectiveExtensibleProgramming2019} to generate performant GEMM code on the fly, including for user-defined elops and for GEMMs on less commonly used data types such as dual numbers, which are useful for automatic differentiation~\cite{dual_numbers}.

Over time, such libraries evolve. For example, cuBLAS has grown to include ever more kernels with fused elops. Ideally, programmers should not have to rewrite their application code to make use of newly available functionality or when switching from one library to another. Instead, they should be able to write generic code that automatically makes use of the available functionality in libraries. Rewrite rules enable this. 

Take for example the expression \code{C .= relu.($\alpha$ * (f\_a.(A) * f\_b.(B))  + $\beta$ * C)}, where \code{A}, \code{B}, and \code{C} are matrices and \code{$\alpha$} and \code{$\beta$} are scalars. The different GEMM and linear algebra packages can each provide a custom set of rewrite rules that rewrite as much as possible of that expression into calls to optimized functions. For example, a rule added to the LinearAlgebra.jl package would allow this expression to be rewritten into a more efficient call to \code{mul!} that carries out the matrix multiplication, scaling by \code{$\alpha$} and \code{$\beta$}, and accumulation into \code{C} at once.
Then only the elops (\code{f\_a}, \code{f\_b}, and \code{relu}) remain to be executed in separate kernels.
For particular argument types, such as 16-bit floating-point matrices in GPU memory, rewrite rules in the GemmKernels.jl package would even allow the complete expression to be fused into one single kernel. 
Other packages could still contain rewrite rules for other GEMM configurations or variants. Depending on which rules are loaded, the same code hence can be rewritten in different ways, exploiting the various and possibly evolving optimization opportunities that complementary libraries offer without having to rewrite the source code.
}

\section{Compilation Time Evaluation}
\label{sec:compiletime}

Julia is a just-ahead-of-time compiled language for scientific computing. By default, Julia automatically compiles a native, type-specialized version of a method just ahead of the first time it is to be executed on some combination of argument types. Julia hence positions itself somewhere between typical ahead-of-time compilers such as LLVM and gcc and just-in-time compilers such as Cranelift or those used in a Java VM. Compilation speed is hence important for Julia. 

A user can also request ahead-of-time compilation of methods and optionally invoke our rewriting rule-based optimization, based on whether or not the investment in compilation time might yield a positive return on investment. As scientific software often has relatively long execution times, compilation times are hence not as critical as in traditional JIT compilers. Still, we aim for interactive compilation speeds with our work to allow for rapid prototyping.

\subsection{Complexity Analysis of Equality Saturation}

\paragraph{E-Graph Construction}
We reimplemented the algorithm that converts IR into an e-graph introduced for Cranelift~\cite{fallinAegraphsAcyclicEgraphs2023} in Julia.
This algorithm starts by computing the dominator tree of the input SSA code.
We reuse the dominator tree construction implementation from the Julia compiler.
This implementation is based on an algorithm from \citeauthor{georgiadisLineartimeAlgorithmsDominators2005}'s thesis~\cite{georgiadisLineartimeAlgorithmsDominators2005}, and runs in linear time with respect to code size.
IR statements are visited in a pre-order dominator tree traversal to ensure that all dependencies of the current statement have already been visited. For each statement, the e-class dependencies are looked up and an e-node is constructed. If the e-node is not part of the e-graph yet, it is inserted in a new e-class, otherwise an existing e-class is associated with the statement.
At the same time that the e-graph is being constructed, the CFG skeleton is being built.
This involves pushing the corresponding e-classes of effectful statements in a CFG structure which has a worst-case execution time complexity of $\mathcal{O}(N)$, where $N$ is the code size, but an amortized complexity of $\mathcal{O}(1)$.

Since no equalities are introduced yet during conversion from IR to e-graph inserting new e-classes in the e-graph does not require running the relatively expensive e-graph rebuilding procedure.
As such, e-graph construction only requires inexpensive hash map lookups and insertions for each statement, which have a worst-case execution time complexity of $\mathcal{O}(N)$ but an expected worst-case execution time of $\mathcal{O}(1)$.

Overall, the worst-case execution time complexity of our e-graph construction is $\mathcal{O}(N^2)$ but on average and in practice, as will be discussed in Section~\ref{sec:experimental compile time}, the worst-case execution time complexity is $\mathcal{O}(N)$.

\paragraph{E-Graph Saturation}
In general, termination for EqSat is not guaranteed because the application of rewrite rules can introduce new rewrite opportunities indefinitely~\cite{willseyEggFastExtensible2021,suciuSemanticFoundationsEquality2025}.
In practice, this issue is most often addressed by EqSat timeouts in combination with heuristic schedulers that ensure that all rewrite rules get a chance to be applied by means of a rule application back-off mechanism~\cite{cheliAutomatedCodeOptimization2021,MLSYS2021_cc427d93}.
The non-terminating behavior of EqSat
is due to repeated rule application and rebuilding the e-graph by joining newly discovered equivalences.

\paragraph{Extraction}
\changed{Constructing the ILP problem entails a few different steps. First, simple cycles are determined using Johnson's algorithm~\cite{johnson1975elementarycircuits}, and the dominator tree is constructed.
Since the number of cycles in a graph can grow exponentially in the number of nodes~\cite{johnson1975elementarycircuits}, in the worst case, the time required to construct the problem is exponential as well.
In practice, however, e-graphs typically contain only a few short simple cycles.
Next, the minimization objective itself can be constructed in $\mathcal{O}(I\cdot N)$, with $N$ the number of e-nodes, and $I$ the number of statements in the CFG skeleton.
Assuming a low number of class cycles, the problem construction time is asymptotically dominated by the construction of the class constraint (Equation~\ref{eq:constraint1}).
In the worst case, this takes 
$\mathcal{O}(I \cdot N \cdot N_{c(max)} \cdot \chi_{\text{(max)}} \cdot D_{\text{(max)}})$ time, with $\chi_{\text{(max)}}$ the maximum number of children e-classes of an e-node, $D_{\text{(max)}}$ the maximum number of dominating statements, and $N_{c(max)}$ the maximum number of e-nodes in an e-class.

Solving the ILP problem is an NP-hard problem, but as will be shown in Section~\ref{sec:compiletime}, solving times are of the same order of magnitude as construction.}

\subsection{Experimental Compilation Time Evaluation}
\label{sec:experimental compile time}
\begin{figure}
    \centering
    \includegraphics[width=\linewidth]{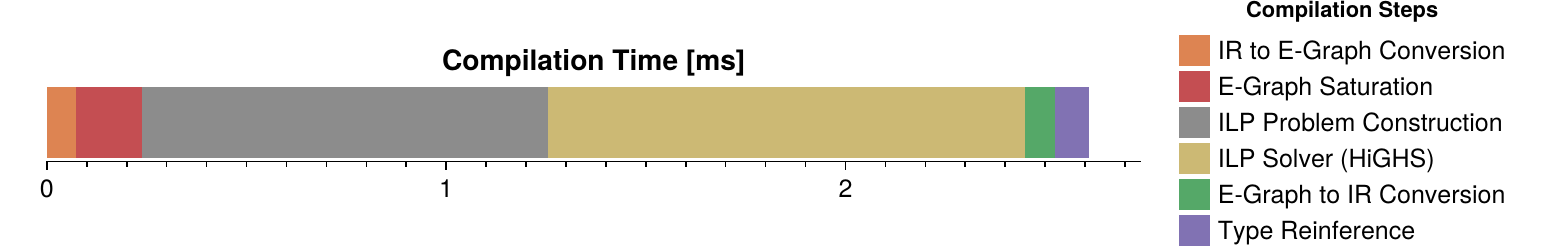}
    \caption{Time spent in different compilation phases for optimizing the \code{forward} method in Fig.~\ref{fig:deep learning original} with the rewrite rules discussed.}
    \label{fig:compilation profile}
\end{figure}

We conducted our measurements on a 64-core AMD Ryzen Threadripper PRO 7985WX and 502GiB RAM, with Julia Version 1.11.5 (LLVM 16.0.6).
Fig.~\ref{fig:compilation profile} shows the time taken for the different steps of our optimization scheme for Fig.~\ref{fig:deep learning original}'s \code{forward} function (invoked with \code{training=false}) and the rewrites to fuse the operations as discussed in Section~\ref{sec:fused}.

\begin{figure}
    \centering
    \begin{minipage}{.58\textwidth}
        \centering
        \includegraphics[height=4.7cm]{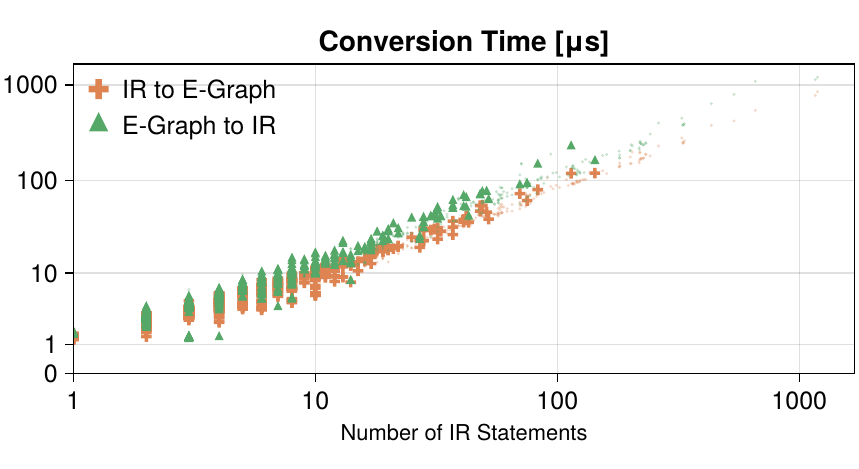}
        \captionof{figure}{Conversion times for converting from IR to e-graph and back. Smaller, faint markers indicate measurements on IR after inlining.}
        \label{fig:conversion profile}
    \end{minipage}\hfill
    \begin{minipage}{.38\textwidth}
        \centering
        \includegraphics[height=4.7cm]{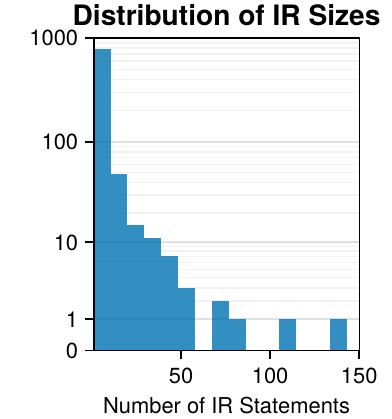}
        \captionof{figure}{A histogram of the number of statements before inlining in IR for typical Julia methods.}
        \label{fig:ir sizes}
    \end{minipage}\end{figure}

\paragraph{Conversion} E-graph conversion, both from and to Julia IR, takes relatively little time for this example.
To evaluate the run time of conversions for other code, Fig.~\ref{fig:conversion profile} shows the time needed to convert between IR and E-Graph for many different Julia methods. These methods were selected arbitrarily by collecting all the methods in the call graphs encountered during the compilation of some methods from Julia's core and linear algebra standard library.
While our optimization targets IR without inlined function calls, we also include timing results for IR with inlined calls to capture performance on longer IR.
Fig.~\ref{fig:ir sizes} shows that the vast majority of the evaluated methods consists of only a handful of IR statements.
This means that we typically do not expect large e-graphs during optimization.

\paragraph{Code Generation} Not shown in the compilation time overview is the time that LLVM spent to optimize and generate the final executable code from Julia IR.
The reason for the omission in the figure is that this step also occurs for regular Julia code compilation.
For the example from Fig.~\ref{fig:compilation profile}, LLVM code generation takes around 75ms, or about 95\% of total compilation time. 
For regular Julia code, much of the compilation cost imposed by LLVM is hidden by caching of compiled code.
Tighter integration of our optimization into the Julia compiler could similarly help hide this cost as well.

\paragraph{ILP Construction and Solving}
\changed{
Fig.~\ref{fig:compilation profile} shows that most of the time is occupied by constructing and solving the ILP extraction problem presented in Section~\ref{sec:extraction}.
Apart from the inherent computational complexity of solving the ILP problem, the conversion takes a large amount of time for converting the e-graph into a collection of variables and constraints in an external ILP solver library.}
Table~\ref{tab:ilp} shows the time required to construct and solve the ILP problem for several use cases discussed in this paper. 
With the exception of the code for simplifying 2D transformations in Fig.~\ref{fig:2D_transformations}, e-classes do not contain many equivalences, leading to fast problem construction and solving.
For the 2D transformation example, particular rewrite rules, e.g., those that combine two subsequent rotations into one rotation, give rise to an infinite number of rewrite opportunities.
In these instances, saturation runs until a timeout is reached.
If fast compilation times are required for pathological cases like the one shown, we can fall back to a greedy extraction algorithm.
In the future, better scheduling of rules and detection of these cases could limit the amount of e-graph blow-up as well.

\begin{table}[t]
    \centering \small
    \begin{tabular}{l|rrcccc}
        \hline
         & \makecell{\textbf{Construction}\\\char`\[ms\char`\]} & \makecell{\textbf{Solve}\\\char`\[ms\char`\]} & \textbf{Variables} & \textbf{Constraints} & \textbf{E-Classes} & \textbf{E-Nodes} \\
        \hline
        Fig.~\ref{fig:extraction} & 0.46 & 1.042 & 126 & 364 & 26 & 28 \\
        Fig.~\ref{fig:2D_transformations} & 24.16 & 208.66 & 918 & 5325 & 337 & 1,002 \\
        Fig.~\ref{fig:deep learning original} & 0.37 & 5.02 & 48 & 168 & 21 & 23 \\
        \hline
    \end{tabular}
    \caption{Time needed to construct and solve the ILP problems for different examples discussed in this work\changed{, and the corresponding number of variables and constraints in the ILP problem, and e-nodes and e-classes in the e-graph.}}
    \label{tab:ilp}
\end{table}

\section{\changed{Performance Evaluation}}
\label{sec:performance}

\begin{table}
    \centering
    \begin{tabular}{lll}
    \textbf{Use Case} &\textbf{Description} & \textbf{Speed-up} \\ \hline
    Fig.~\ref{fig:2D_transformations} &2D transformations & $16.5\times$       \\
    Fig.~\ref{fig:extraction} &Linear algebra value reuse & up to $1.1\times$      \\
    Fig.~\ref{fig:deep learning original} &Fused neural network & up to $2.5\times$ \\
    Sec.~\ref{sec:gemm} &GEMM rewrites & up to $4.8\times$
    \end{tabular}
    \caption{Speed-up for the optimized code compared to the unoptimized code for various use cases.}
    \label{tab:speedups}
\end{table}
 \begin{figure}   
\includegraphics[width=0.47\textwidth]{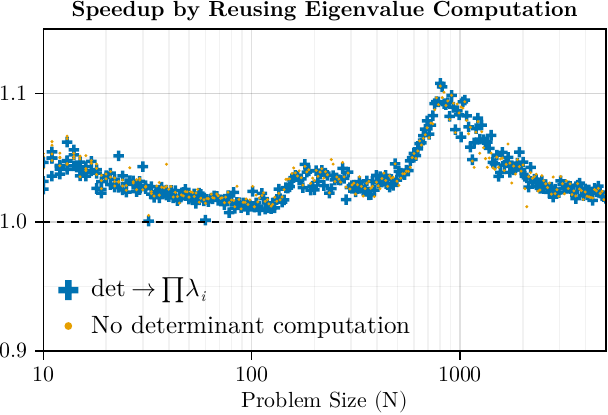}
    \caption{Speed-up over the original code of Fig.~\ref{fig:extraction}. The problem size $N$ refers to the 64-bit floating point input matrix of size $N \times N$. Computation was run on the CPU.}
    \label{fig:speedup eigen}
\end{figure}

 Our work does not at all aim to squeeze more performance out of the Julia programming language and its compiler and run-time implementation. Indeed, any optimization that can be obtained with a rewrite rule can also be obtained by manually rewriting source code, such as by duplicating source code and then special-casing it. Still, as the goal of the rewrite rules is to enable optimizations, this section reports the performance gains that can be achieved for some of the previously presented use cases.

Our results are summarized in Table~\ref{tab:speedups}. The use case of Fig.~\ref{fig:2D_transformations} where domain-specific code representing a 2D transformation is optimized into an idempotent function shows a large speed-up, as expected by all computations being optimized out of the function.
The other examples paint a more interesting picture.

For the use case of Fig.~\ref{fig:extraction}, where the computation of a matrix determinant is simplified by reusing the already computed eigenvalues, Fig.~\ref{fig:speedup eigen} shows a moderate speed-up across the board. For problem sizes around $N=1000$, this speed-up is more pronounced.
For reference, yellow dots show the speed-up when simply omitting the determinant computation completely. In other words, they show what speed-up can be achieved by reducing the execution time of the determinant computation to zero. The fact that those roughly match with the optimized code shows that in the rewritten code, the determinant value basically comes for free, whereas it costs up to 10\% in the original code.

\begin{figure}
    \centering
\begin{minipage}[t]{.47\textwidth}
        \centering
        \includegraphics[width=\linewidth]{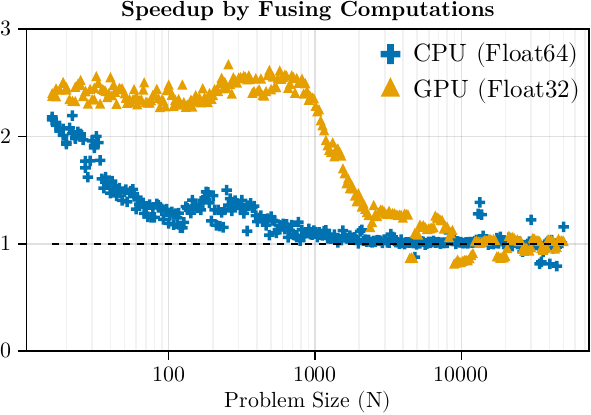}
        \caption{Speed-up gained over the original code of Fig.~\ref{fig:deep learning original}. The problem size $N$ corresponds to the size of the matrices used in the computation:\quad
        \raisebox{-0.35cm}{\includegraphics[width=0.35\linewidth]{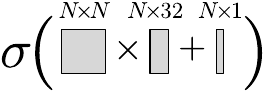}}}
        \label{fig:speedup fused}
    \end{minipage}\hfill
    \begin{minipage}[t]{.47\textwidth}
        \centering
        \includegraphics[width=\linewidth]{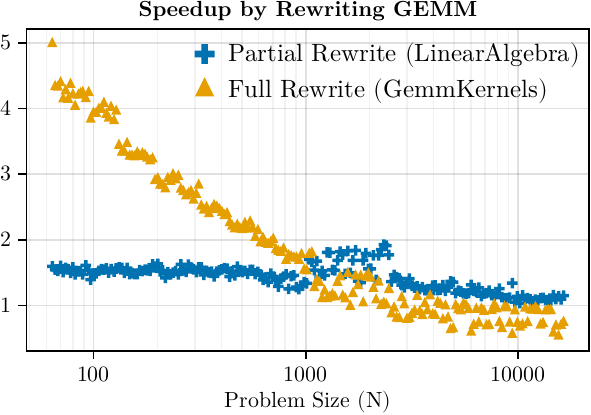}
        \caption{Speed-up when rewriting the GEMM expression from Sec.~\ref{sec:gemm} into a combination of elops and an optimized \code{mul!} operation (blue), and a single fused GemmKernels.jl kernel (yellow). The problem size $N$ refers to the 16-bit floating point input matrices of size $N \times N$. All computations on GPU.}
        \label{fig:speedup gemm}
    \end{minipage}
\end{figure}

Similarly, for the use case of Fig.~\ref{fig:deep learning original} where the forward pass of a neural network is replaced with a fused implementation, the plot in Fig.~\ref{fig:speedup fused} shows a speed-up for small problem sizes but ends up about as fast as the original code for larger problem sizes, as the time spent in the matrix multiplication tends to dominate.~\footnote{\changed{For problems involving GPU computation (Fig~\ref{fig:speedup fused}, Fig.~\ref{fig:speedup gemm}), we measured execution time on a machine with a 16-core Intel Xeon E5-2637 v2, NVIDIA GeForce RTX 2080 Ti (12GB), and 64GB RAM, with Julia Version 1.11.5 (LLVM 16.0.6). For the other problems, we use the same setup as in Section~\ref{sec:compiletime}.}} 

\changed{
Finally, we look at the example of rewriting the expression \code{C .= relu.($\alpha$ * (f\_a.(A) * f\_b.(B))  + $\beta$ * C)}, discussed in Section~\ref{sec:gemm}. Fig.~\ref{fig:speedup gemm} shows that the basic linear algebra rewrite, which fuses everything except the elops (\code{f\_a}, \code{f\_b}, \code{relu}) into one GPU kernel. This provides a moderate speed-up across the board over the baseline implementation in which each operation in the whole expression corresponds to a separate GPU kernel. The GemmKernels rewrite, which fuses all operations into a single kernel, is worthwhile for smaller problem sizes, and shows a small slowdown for larger problems, where the core GEMM computation dominates.

}

\changed{This performance evaluation of four uses cases demonstrates that our system can indeed be used to achieve significant speed-ups by exploiting domain-specific optimization opportunities, in a programmer-friendly manner.}

\section{Limitations}
\label{sec:limitations}
\paragraph{Control Flow and Inlining}
As discussed in \changed{Section~\ref{sec:rewriter}}, our approach to converting IR to an e-graph using the CFG skeleton currently does not support rewrites that alter control flow.
Similarly\changed{, since we apply rewrites on code where function calls have not yet been inlined}, rewrite opportunities where part of an expression is hidden behind a function call are not considered.
We optimize each method separately.
Other intermediate representations, discussed in Section~\ref{sec:related work}, could be used to represent the entire program in an e-graph and allow rewriting across call boundaries and control flow. 

\paragraph{Saturation and Extraction}
In this work, we have proposed an ILP formulation for optimal e-graph extraction.
The problem of extracting e-nodes from a general e-graph is NP-hard~\cite{goharshadyFastOptimalExtraction2024}, which means that there is no guarantee that an optimal solution can be found \changed{in polynomial time}.
Currently, when the ILP solver is unable to find a solution \changed{within a given timeframe}, we fall back to a greedy extraction algorithm that is built into Metatheory.jl.
In this work, we chose a simple cost model that treats the cost of nodes independent of other nodes.
A more complex cost model could take into account more complex behavior that depends on multiple nodes at once. 

\paragraph{Rule Expressiveness}
As discussed in Section~\ref{sec:rewriter}, our rule syntax takes one Julia expression and transforms it into another.
This is an intuitive syntax, but does not fully cover all the possible rules a developer could possibly want to express.
For one, the rewrite rules can only be used to rewrite a \textit{single} expression into another expression.
Multi-pattern rewrite rules to match against and to rewrite multiple expressions at once are currently not supported.
We also do not yet support rewrite rules matching against function calls taking an unknown, variable amount of variables. A developer may hence have to rewrite multiple similar rules for the same function with different argument counts.

\changed{
\paragraph{Rule applicability}
Section~\ref{sec:performance} shows that certain rewrites do not always yield positive returns. Consequently, the burden is currently on the programmer to select the rule sets for their particular problem.
In the future, this could be automated using an improved cost model, in combination with more detailed program analysis to gauge the effectiveness of rewrite rules.
Alternatively, the rule applicability could be made more extensible, allowing package developers to decide programmatically when a rule can be applied.
}

\section{Related Work}
\label{sec:related work}
\subsection{Equality Saturation}
EqSat and its use as a code optimization tool was first introduced by \citeauthor{tateEqualitySaturationNew2009}~\cite{tateEqualitySaturationNew2009}, where it was used to optimize Java bytecode.
\citeauthor{willseyEggFastExtensible2021}~\cite{willseyEggFastExtensible2021} introduced algorithmic improvements for EqSat providing asymptotic speed-ups.
This has led to an explosion of new work exploring different applications for EqSat, from optimizing tensor programs~\cite{wangSPORESSumproductOptimization2020,MLSYS2021_cc427d93,smithPureTensorProgram2021}, to compiler verification~\cite{steppEqualityBasedTranslationValidator2011,kourtaCaviarEgraphBased2022}, optimizations for specialized hardware~\cite{ustunIMpressLargeInteger2022,vanhattumVectorizationDigitalSignal2021,matsumuraSymbolicEmulatorShuffle2023}, and other purposes~\cite{panchekhaAutomaticallyImprovingAccuracy2015,chandrakananandiRewriteRuleInference2021}.

Metatheory.jl~\cite{cheliMetatheoryjlFastElegant2021} provides a framework for EqSat in Julia.
It has already been used for code optimization on symbolic representations of Julia code to speed up PDE solvers~\cite{gowdaHighperformanceSymbolicnumericsMultiple2022}. That work, however, only handles pure, symbolic programs, as it operates on a symbolic representation of Julia code obtained through manual expression building or tracing through heavily restricted, straight-line code. In contrast to our work, the existing use of Metatheory.jl does hence not allow the presence of general control flow or side effects.

To the best of our knowledge, by introducing type propagation in the e-graph, we are the first to apply EqSat to a dynamic programming language.
This allows rewrites to coexist and complement Julia's multiple dispatch feature, as discussed in Section~\ref{sec:usecase}.

\subsection{Intermediate Representations and E-Graphs}
Cranelift, a compiler back-end for WebAssembly and Rust, is among the first to use e-graphs and EqSat in a production-grade compiler~\cite{fallinAegraphsAcyclicEgraphs2023}. While we reuse ideas and methods first introduced in the Cranelift compiler, we implemented our framework as a standalone project, not reusing any Cranelift components. Our code is integrated in the Julia compiler using user-land compiler extensions.

The Cranelift compiler does not do full EqSat, but only recognizes equalities the first time a node is inserted in the e-graph.
This implies that it is possible for equalities to remain undiscovered, but it obviates the fixed-point loop present in full EqSat and ensures there are no cycles in the e-graph which allows for a more efficient representation of the graph in memory and leads to easier extraction.
We use the same e-graph representation of code as used in Cranelift, but by utilizing an extraction scheme that enforces acyclic extraction, the optimizer is able to carry out full EqSat.
Whereas Cranelift and our work operate on IR that is complemented by a CFG, other work has sought to find alternative IR formats that allow representing control flow fully in the e-graph.
\citeauthor{tateEqualitySaturationNew2009}~\cite{tateEqualitySaturationNew2009} have introduced Program Expression Graphs (PEGs) which represent constructs such as loops as special expressions and show that PEGs can be used to do EqSat.
\changed{\citeauthor{LairdSpEQ2024}~\cite{LairdSpEQ2024} raise supported LLVM IR into a functional IR that also allows control flow transformations.}
Similarly, Regionalized Value State Dependency Graphs (RVSDGs)~\cite{bahmannPerfectReconstructabilityControl2015,reissmannRVSDGIntermediateRepresentation2020} represent control flow as (nested) expressions.
A research prototype already exists using RVSDGs in conjunction with EqSat for a simple toy language~\cite{optir2022}.
Our approach remains closer to the original SSA-based IR of the source language, allowing less expensive conversion routines.
\changed{Lastly, there are multiple works applying equality saturation on MLIR code~\cite{merckx2025eqsatequalitysaturationdialect,zayed2025dialegg}. 
These do not keep track of side effects and control flow in a CFG skeleton and thus can only operate on code that is pure. Additionally, they operate on statically typed IR in contrast to our work.
}

\citeauthor{willseyEggFastExtensible2021} introduced e-class analyses\changed{, a framework} to associate additional metadata with each e-class~\cite{willseyEggFastExtensible2021}. The metadata is kept up to date throughout the EqSat process by potential modifications whenever new information is available, for example, when e-classes are merged. An exemplary existing use case for an e-class analysis is constant propagation, where each e-class can optionally carry a constant value.
\changed{Further work has shown that this technique can be combined with abstract interpretation for more complex program analyses~\cite{cowardCombiningEGraphsAbstract2023}.}
To track types in the e-graph, we designed a novel e-class analysis, as discussed in Section~\ref{sec:types}.

\subsection{Extraction}
We based our ILP optimal extraction formulation on the work of \citeauthor{heImprovingTermExtraction2023}~\cite{heImprovingTermExtraction2023}.
Recent work proposes other formulations, for example by looking at e-graphs through the lens of circuits~\cite{sunEgraphsCircuitsOptimal2024}, or finite state automata~\cite{y.wangEGraphsVSAsTree2022}.
These formulations can lead to faster extraction or provide termination guarantees under particular assumptions.
Cranelift operates in a just-in-time context where fast and reliable extraction is crucial.
For this reason, they use an extraction algorithm that greedily tries to minimize the total number of nodes, without taking into account node reuse in extracted expressions or from expressions that dominate them.
In contrast to Cranelift's greedy algorithm, we have introduced an ILP formulation that aims for node reuse from other nodes within an extracted expression, as well as from nodes in expressions that dominate them in the CFG. 

\subsection{Julia Code Optimization}
Other works exist that aim to improve the performance of Julia code with custom optimizations.
Symbolics.jl~\cite{gowdaHighperformanceSymbolicnumericsMultiple2022} can be used to transform scientific machine learning code into a symbolic representation on which simplifications can be applied.
Finch.jl~\cite{ahrensFinchSparseStructured2024,ahrensLoopletsLanguageStructured2023} uses a custom IR to represent iteration patterns of loop nests over sparse or structured arrays.
We instead focus on optimizing code in Julia's own IR format, allowing users of our system to target any Julia code.

\section{Conclusion and Future Work}
\label{sec:conclusions}
We have developed a novel system that allows Julia developers to write domain-specific rewrite rules that are automatically applied using equality saturation. Our system works in the presence of control flow and side effects.
Through an e-class analysis, we keep track of the most specific type of all equivalent terms and use this information to support type-constrained rewrite rules.
Unlike previous work, our system can rewrite dynamically-typed code.
We introduce a technique called CFG skeleton relaxation that allows rewrite rules to nullify side effects in the original code.
We have adapted an ILP formulation for optimal, acyclic e-graph extraction to take into account value reuse from dominating statements.
Finally, we show how this system can perform rewrites on high-level domain-specific code and how the use of equality saturation solves the phase-ordering problem.

In the future, our work can be extended to support rewriting control flow and apply rewrites that span different call depths\changed{, by integrating function call inlining in the equality saturation procedure}.
Further improvements to our extraction scheme could also lead to more efficient code being generated.

All artifacts will be made available upon acceptance of the paper. \newpage

\bibliographystyle{ACM-Reference-Format}
\bibliography{references.bib}

\end{document}